\documentstyle[preprint,prl,aps]{revtex}
\input psfig
\begin{document}
\preprint{LBNL-40398, June, 1997}

\title{Charm Quark Production in Non-central Heavy Ion Collisions}
\author{V. Emel'yanov$^1$, A. Khodinov$^1$, S. R. Klein$^2$, R. Vogt$^{2,3}$}

\address{{
$^1$Moscow State Engineering Physics Institute (Technical
University), Kashirskoe ave. 31, Moscow, 115409, Russia\break 
$^2$Nuclear Science Division, Lawrence Berkeley National Laboratory, 
Berkeley, CA
94720, USA\break 
$^3$Physics Department, University of California, Davis, CA 95616, USA}\break}

\vskip .25 in
\maketitle
\begin{abstract}

The effect of gluon shadowing on charm quark
production in large impact parameter
ultrarelativistic heavy ion collisions is investigated.
Charm production cross sections are calculated for a range of non-central
impact parameters which can be accurately inferred from the global
transverse energy distribution.  We show that charm production is a good
probe of the local parton density which determines the effectiveness of
shadowing. The spatial dependence of shadowing can only be studied in heavy ion
collisions.
\vskip .5 in
\centerline{(Submitted to Physical Review C.)}

\end{abstract}
\pacs{1}

\section{Introduction}

Deep inelastic scattering experiments using nuclear targets 
showed that the quark and antiquark distribution functions
are modified in the nuclear environment \cite{Arn} and hence are different
in heavy nuclei than in free protons.  It is not unreasonable to
expect the nuclear gluon distributions to be affected at least as much
as the quark distributions.  However, little is known about the nuclear gluon 
distribution because the gluon distributions can only be indirectly probed.
Gluon-dominated production processes, such as $J/\psi$ and heavy quark
production, can provide an indirect measure of the nuclear gluon distribution.
Since the $J/\psi$ is more strongly affected by absorption processes than charm
quarks, evident from their respective $A$ dependencies \cite{Alde,Appel}, 
charmed  quark production
provides a cleaner determination of the nuclear gluon distribution.

To date, all measurements  and indirect determinations of nuclear parton
distributions have been insensitive to the position of the interacting parton
within the nucleus.  However, there is no reason to expect the parton momentum
distributions to be constant within the nucleus. They should at least
vary with the local nuclear density.  If shadowing is due to gluon 
recombination, the position dependence could be quite strong \cite{hot}.
One way to probe the position dependence of the shadowing is to measure 
$c \overline c$ production over a wide range of
impact parameters, thus scanning gluon localization in the nucleus.  The charm
rate has been shown to be large in central collisions
\cite{GMRV}, here we will show that these studies are also feasible at large
impact parameters.

This paper thus proposes a method for measuring the position dependence of the
gluon momentum distribution in heavy nuclei.  We show that the charm
production rates in non-central collisions are sensitive to the
details of the gluon distribution and its position dependence.  We use
two different parameterizations of nuclear shadowing along with two
parameterizations of the position dependence of the shadowing to
calculate charm production in 100 GeV per nucleon Au+Au collisions at
the Relativistic Heavy Ion Collider (RHIC)\cite{RHIC}, now under
construction at Brookhaven National Laboratory.  However, the
techniques discussed here should also be applicable to $c\overline c$
and $b\overline b$ production in Pb+Pb collisions at the CERN Large
Hadron Collider (LHC).  The charm quark production rate and $p_T$
spectra are calculated as a function of impact parameter, $b$, for
non-central collisions with impact parameters greater than the nuclear
radius, $R_A$.

For this study, we need to select events according to impact parameter.
Although the impact parameter of the collision is not directly measurable,
it may be inferred from the total transverse energy, $E_T$, of the
event\cite{us}.  We discuss the relationship between $E_T$ and
$b$ and present calculations showing that, for a given
$E_T$, the impact parameter can be measured relatively accurately.  Additional
input, such as a measurement of nuclear breakup, through the use of a zero 
degree calorimeter, can refine this estimate.

\def\ccbar{$c\overline c$} 

Section 2 summarizes the calculations of \ccbar\ production in peripheral 
collisions including a discussion of the nuclear parton shadowing and its
possible spatial dependence.
Section 3 discusses the relationship between transverse energy and
impact parameter.  Section 4 presents the numerical results for
the charm production rates and $p_T$ spectra for two ranges
of non-central impact parameters.  We demonstrate how these rates are 
sensitive to the nuclear gluon distribution.  Our 
results are put into an experimental perspective
in Section 5.  Finally, Section 6 draws some conclusions.

\section{$c\overline c $ Production}

\newcommand{\be}{\begin{eqnarray}}
\newcommand{\ee}{\end{eqnarray}}
\renewcommand{\thefootnote}{\alph{footnote}}

To study the effects of shadowing on $c \overline c$ production in
peripheral collisions, we emphasize the modifications of the parton
distribution functions due to shadowing as well as the location of the
interacting parton in the nucleus.  We discuss the method used to
calculate $c \overline c$ pair production and introduce two
parameterizations of nuclear shadowing.  We also describe two models
of the spatial dependence of the shadowing.

The double differential cross section
for $c \overline c$ pair production by nuclei $A$ and $B$ 
is \be
\lefteqn{ E_c E_{\overline c} \frac{d\sigma_{AB}}{d^3p_c d^3p_{\overline c} 
d^2b d^2r} = } \\ & & \sum_{i,j}\int \,dz \,dz'
\,dx_1\, dx_2 F_i^A(x_1,Q^2,\vec{r},z) F_j^B(x_2,Q^2,\vec{b} - \vec{r},z')
E_c E_{\overline c} \frac{d\widehat{\sigma}_{ij}
(x_1P_1,x_2P_2,m_c,Q^2)}{d^3p_c d^3p_{\overline c}} \nonumber \, \,  . \ee
Here $i$ and $j$ are the interacting partons in the nucleus
and the functions $F_i$ are the number densities of gluons, light 
quarks and antiquarks
evaluated at momentum fraction $x$, scale $Q^2$, and location $\vec{r}$, $z$.
(Note that $\vec{r}$ is two-dimensional.)   
The short-distance cross section,
$\widehat{\sigma}_{ij}$, is calculable as a perturbation series in
$\alpha_s(Q^2)$.

At leading order (LO),
$c \overline c$ production proceeds by two basic processes,
\be
 q + \overline q  & \rightarrow & c + \overline c \\
 g + g & \rightarrow  & c + \overline c  \, \, .
\ee
The LO cross section, ${\cal O}(\alpha_s^2)$, can be written as
\be 
\lefteqn{E_c E_{\overline c} \frac{d\sigma_{AB}}{d^3p_c d^3p_{\overline c} 
d^2b d^2r} = } \\
& & \int \frac{s}{2 \pi} \,dz \, dz' \,dx_1\, dx_2 \, 
C(x_1,x_2,Q^2,\vec{r},z,\vec{b}- \vec{r}, z') \, \delta^4(x_1P_1
+ x_2P_2 - p_c - p_{\overline c}) \nonumber \, \, \ee
where $\sqrt{s}$, the
parton-parton center-of-mass energy, is related to $\sqrt{S}$, the
hadron-hadron center-of-mass energy, by $s = x_1 x_2 S \geq 4m_c^2$,
where the momentum fractions, $x_1$ and $x_2$, are
\be x_{1,2} = \frac{m_T}{\sqrt{s}} (e^{\pm y} + e^{\pm \overline y})
\, \, , \ee and $m_T = \sqrt{m_c^2 + p_T^2}$.  
The target fraction, $x_2$, decreases with rapidity while the
projectile fraction, $x_1$, increases.
Here, the intrinsic transverse 
momenta of the incoming partons has been neglected.
The convolution of the subprocess
cross sections with the parton number densities is contained in
$C(x_1,x_2,Q^2,\vec{r},z,\vec{b}- \vec{r},z')$ where \be 
\lefteqn{C(x_1,x_2,Q^2,\vec{r},z,\vec{b}
- \vec{r},z')  = } \\ & & \sum_q
[F_q^A(x_1,Q^2,\vec{r},z)  F_{\overline q}^B(x_2,Q^2,\vec{b} - \vec{r},z') +  
F_{\overline q}^A(x_1,Q^2,\vec{r},z) F_q^B(x_2,Q^2,\vec{b} - \vec{r},z')]
\frac{d
\widehat{\sigma}_{q \overline q}}{dt} \nonumber \\
& & + F_g^A(x_1,Q^2,\vec{r},z) F_g^B(x_2,Q^2,
\vec{b} - \vec{r},z') 
\frac{d \widehat{\sigma}_{gg}}{dt} \nonumber
\, \, . \ee
Four-momentum conservation leads to the rather simple expression
\be \frac{d \sigma_{AB}}{dp_T^2 dy d\overline y d^2b d^2r} = \int dz \,dz'\, 
x_1 x_2 C(x_1,x_2,Q^2,\vec{r},z,\vec{b} - \vec{r},z') \, \, , \ee
The LO subprocess cross sections for $c \overline c$ production by
$q\overline q$ annihilation and $gg$ fusion,
expressed as a function of
$m_T$, $y$, and $\overline y$, are \cite{Ellis}
\be
\frac{d \hat{\sigma}_{q \overline q}}{dt} = \frac{\pi
\alpha_s^2}{9 m_T^4} \frac{\cosh(y - \overline y) +
m_c^2/m_T^2}{(1+ \cosh(y - \overline y))^3} \, \, ,
\ee
\be  \frac{d \hat{\sigma}_{gg}}{dt} = \frac{\pi \alpha_s^2}{96
m_T^4} \frac{8 \cosh(y - \overline y) - 1}{(1+
\cosh(y - \overline y))^3} \left(
\cosh(y - \overline y) + \frac{2m_c^2}{m_T^2} -
\frac{2m_c^4}{m_T^4} \right) \, \, .
\ee
Leading order calculations tend to underestimate the measured charm production
cross section by a constant factor, usually called a $K$ factor,
\begin{equation} 
K_{\rm exp}^{\rm LO} = \frac{\sigma_{\rm
exp}(AB \rightarrow c \overline c)}{\sigma_{\rm LO}(AB \rightarrow c
\overline c)} \, \, , 
\end{equation} 
The next-to-leading order (NLO) corrections to the LO cross section
have been calculated \cite{NDE,SvN} and an analogous theoretical $K$
factor $K_{\rm th}$ can be defined from the ratio of the NLO to the LO
cross sections, 
\begin{equation} 
K_{\rm th} = \frac{\sigma_{\rm
NLO}(AB \rightarrow c \overline c)}{\sigma_{\rm LO}(AB \rightarrow c
\overline c)} \, \, , 
\end{equation} 
where $\sigma_{\rm NLO}$ is the
sum of the LO cross section and the ${\cal O}(\alpha_s)$ corrections.

Previously \cite{HPC}, the NLO calculations were compared to the $c
\overline c$ total production cross section data to fix $m_c$ and $Q$
so that $K_{\rm exp}^{\rm NLO} \sim 1$ to provide a more reliable estimate
for nuclear collider energies.  Reasonable
agreement with the measured total cross section was found for $m_c =
1.2$ GeV, $Q = 2m_c$ for MRS D$-^\prime$ \cite{MRS} and $m_c = 1.3$
GeV, $Q = m_c$ for GRV HO \cite{GRV}.  We choose different scales for the two
sets\footnote{These structure functions can be found in the CERN 
PDFLIB \cite{pdflib}.} because of the different initial scales of the two
parton distributions.
The MRS D$-^\prime$ distributions have $Q_{0, {\rm MRS}}^2 = 5$ GeV$^2$; we
choose $Q = 2m_c$ so that $Q^2 > Q_{0, {\rm MRS}}^2$.  The GRV HO sea quark and
gluon distributions are valence-like at low $x$ and $Q_{0, {\rm GRV}}^2 = 0.3$
GeV$^2$.  We can then use $Q = m_c$ because $m_c^2 > Q_{0, {\rm GRV}}^2$.
However, below $Q^2 \approx 5$ GeV$^2$ the gluon distribution is still somewhat
valence-like.

When calculating inclusive distributions rather than total cross
sections, it is more appropriate to choose $Q \propto m_T$,
particularly when $p_T > m_c$ since a constant scale introduces
unregulated collinear divergences \cite{rvz}.  Therefore, we take $Q =
2m_T$ for the MRS D$-^\prime$ distributions and $Q = m_T$ for the GRV
HO distributions.  Both sets of parton densities result in a NLO total
$c \overline c$ production cross section of $\sim 350$ $\mu$b in $pp$
collisions at $\sqrt{s} = 200$ GeV.

The differential $K_{\rm th}$ for the charm quark $p_T$ distribution,
the pair mass distribution, and the charm quark and $c \overline c$ pair
rapidity distributions are nearly constant at RHIC energy\cite{rvz}.
They are also essentially independent of the parton density.  The
value of $K_{\rm th}$ is determined by a comparison of the NLO
and LO total cross sections.  Our LO calculations, eq.\ (4), are
multiplied by the appropriate $K_{\rm th}$ found for the specific
parton density: 2.5 for the MRS D$-^\prime$ distributions and 2.9 for
the GRV HO distributions.

The nucleon parton densities are only a part of the space-dependent nuclear
number
densities, $ F_i^A(x,Q^2,\vec{r},z)$, introduced in eq.\ (1).  We have assumed
that these nuclear number densities factorize into nuclear density 
distributions,
independent of $x$ and $Q^2$, the nucleon parton densities, independent of
spatial position and $A$, and a shadowing function that parameterizes the 
modifications of the nucleon parton densities
in the nucleus, dependent on $x$, $Q^2$, $A$ and location,
\be
F_i^A(x,Q^2,\vec{r},z) & = & \rho_A(s) S^i(A,x,Q^2,\vec{r},z) 
f_i^p(x,Q^2) \\
F_i^B(x,Q^2,\vec{b} - \vec{r},z') & = & \rho_B(s') S^i(B,x,Q^2,\vec{b} - 
\vec{r},z') f_i^p(x,Q^2) \nonumber \,\, ,
\ee
where $s = \sqrt{r^2 + z^2}$, 
$s'=\sqrt{|\vec{b}-\vec{r}|^2 +z'^{\, 2}}$ 
and $f_i^p$ are the nucleon parton densities. We assume that $z$ and $z'$
are uncorrelated.  The collision geometry in the plane transverse to the
beam is shown in Fig.\ 1.

A three parameter Wood--Saxon shape is used to describe the nuclear density
distribution, \begin{eqnarray} \rho_A(s)=\rho_0 \frac{1 + \omega(s/R_A)^2}{1 +
\exp((s-R_A)/d)} \, \, , \end{eqnarray} where $R_A$ is the nuclear radius,
$d$ is the surface thickness, and $\omega$ allows for central irregularities. 
The electron scattering data of Ref.\ \cite{Vvv} is used
for $R_A$, $d$, and $\omega$ assuming that the charge and matter density 
distributions are identical. The central density, $\rho_0$, is found from 
the normalization $\int d^2r dz \rho_A(s) = A$. For gold,
$\omega = 0$, $d = 0.535$ fm, $R_A = 6.38$ fm, and $\rho_0 = 0.1693$ fm$^{-3}$.

If the parton densities in the nucleon and in
the nucleus are the same, then $S^i(A,x,Q^2,\vec{r},z) \equiv 1$.  
We will use this as a baseline
against which to compare our results with shadowing included.  

We now discuss our choices of the shadowing parameterizations used in our
calculations, independent of the position.
Measurements of the nuclear 
charged parton distributions by deep-inelastic scattering
on a nuclear target and a deuterium target, show that the ratio $R_{F_2} =
F_2^A/F_2^D$
has a characteristic shape as a function of $x$.  The region  
below $x\sim 0.1$ is referred to as the shadowing region and the region
$0.3<x< 0.7$ is known as the EMC region.  
In both regions a depletion is observed in
the heavy nucleus relative to deuterium and $R_{F_2}<1$.  
At very low $x$, $x \approx 0.001$, $R_{F_2}$ appears to
saturate\footnote{We note that at
even smaller values of $x$, shadowing within the nucleon itself is expected
\cite{hot,GLRMQ}.  However, at RHIC energies, this very low $x$ region is not
expected to be reached.} \cite{E6652}.  Between the shadowing and EMC 
regions, an enhancement, antishadowing, occurs where
$R_{F_2}>1$.  There is also an enhancement as $x \rightarrow 1$, assumed to be
due to Fermi motion of the nucleons.  The general behavior of $R_{F_2}$ 
as a function of $x$ is often
referred to as shadowing.  Although this behavior is 
not well understood for all $x$, 
the shadowing effect can be modeled by an $A$ dependent fit to the nuclear
deep-inelastic scattering data and implemented by a modification of the parton
distributions in the proton.  We use two different models of 
the relation between $R_{F_2}$ and $S^i(A,x,Q^2)$.
These two parameterizations were used earlier to estimate
the effect of shadowing on $c \overline c$ and $b \overline b$
production in central collisions \cite{GMRV} with no spatial dependence assumed
for the shadowing.

The first parameterization
is a fit to recent nuclear 
deep-inelastic scattering
data.  The fit does not differentiate between quark, antiquark, and gluon
modifications and does not include evolution in $Q^2$.  Therefore
it is not designed to conserve baryon number or momentum.  We define
$R_{F_2} = S_1(A,x)$ \cite{EQC} with \be S_1(A,x) = \left\{
\begin{array}{ll} R_s {\displaystyle \frac{1 + 0.0134 (1/x -1/x_{\rm sh})}{1
+ 0.0127A^{0.1} (1/x - 1/x_{\rm sh})}} & \mbox{$x<x_{\rm sh}$} \\
a_{\rm emc} - b_{\rm emc}x  & \mbox{$x_{\rm sh} <x< x_{\rm fermi}$} \\
R_f \bigg( {\displaystyle \frac{1-x_{\rm fermi}}{1-x}} \bigg)^{0.321} &
\mbox{$x_{\rm fermi} <x< 1$} \end{array} \right. \, \, , \ee
where $R_s = a_{\rm emc} - b_{\rm emc} x_{\rm sh}$, $R_f = a_{\rm emc} -
b_{\rm emc} x_{\rm fermi}$, $b_{\rm emc} = 0.525(1 - A^{-1/3} - 1.145A^{-2/3} +
0.93A^{-1} + 0.88A^{-4/3} - 0.59A^{-5/3})$, and $a_{\rm emc} = 1 + b_{\rm emc}
x_{\rm emc}$.   The fit fixes $x_{\rm sh}=0.15$, 
$x_{\rm emc}=0.275$ and $x_{\rm fermi}=0.742$.
Thus, the nuclear parton densities are modified so
that \be f_i^A(x,Q^2) & = & S_1(A,x) f_i^p(x,Q^2) . \ee

The second parameterization, $S_2^i(A,x,Q^2)$, modifies the
valence and sea quark and gluon distributions separately and also includes 
$Q^2$ evolution\cite{KJE}, 
but is based on an older fit to the data using the
Duke-Owens parton densities \cite{DO}.  
The initial scale for the evolution is $Q_0 = 2$ GeV and the $Q^2$
evolution is studied with both the standard Altarelli-Parisi evolution and with
modifications due to gluon recombination  
at high density.  The gluon recombination terms do not
strongly alter the evolution.
In this case, the nuclear parton densities are modified so
that \be f_V^A(x,Q^2) & = & S_2^V(A,x,Q^2) f_V^p(x,Q^2) \\  
f_S^A(x,Q^2) & = & S_2^S(A,x,Q^2) f_S^p(x,Q^2) \\ f_G^A(x,Q^2) & =
& S_2^G(A,x,Q^2) f_G^p(x,Q^2) \, \, , \ee where $f_V = u_V + d_V$ is 
the valence 
quark density and $f_S = 2(\overline u + \overline d + \overline s)$ is
the total
sea quark density.  We assume that $S_2^V$ and $S_2^S$ affect the up, down, and
strange valence and sea
quarks identically.  The ratios were constrained by
assuming that $R_{F_2} \approx S_2^V$ at large $x$ and 
$R_{F_2} \approx S_2^S$ at
small $x$ since $xf_V^p(x,Q_0^2) \rightarrow 0$ as $x \rightarrow 0$.  
For the gluons, we take $R_{F_2} \approx S_2^G$ for all $x$ \cite{KJE},
since one might expect more shadowing for the sea quarks, generated from
gluons, at small $x$.
These parton densities do conserve baryon number,
$\int_0^1 dx \, f_V^{p,A}(x,Q^2) = 3$, and
momentum,
$\int_0^1 dx \, x(f_V^{p,A}(x,Q^2) + f_S^{p,A}(x,Q^2) + f_G^{p,A}(x,Q^2)) = 1$.
at all $Q^2$.  We have used the MRS D$-^\prime$ and GRV HO densities with 
$S_2^i$ instead of the Duke-Owens densities, leading to some
small deviations in the momentum sum but the general trend is unchanged.

Since the shadowing is likely related to the nuclear density, it
should also depend on the spatial distribution of the partons within
the nucleus so that $S^i(A,x,Q^2,\vec{r},z) \rightarrow 1$ as $s \rightarrow
\infty$.  The reduced shadowing is reasonable since the shadowing
mechanism should be less effective when the nuclear density is low.
This spatial dependence should also be normalized so that $\frac{1}{A}
\int d^2r dz \rho(s) S^i(A,x,Q^2,\vec{r},z) = S^i(A,x,Q^2)$ to recover
the deep-inelastic scattering results which do not have any explicit
impact parameter dependence.  This approach may fail when
$x\rightarrow 1$, because then the change in the structure function
is likely due to Fermi motion, which should not exhibit spatial
dependence.  

One natural parameterization of the spatial dependence follows the nuclear
matter density distribution,
\be
S^i_{\rm WS} = 
S^i(A,x,Q^2,\vec{r},z) & = & 1 + N_{\rm WS}
\frac{S^i(A,x,Q^2) - 1}{1 + \exp((s-R_A)/d)} \\
                     & = & 1 + N_{\rm WS}
(S^i(A,x,Q^2) - 1) \frac{\rho(s)}{\rho_0}
                     \nonumber \, \, , 
\ee
where $N_{\rm WS} = 1.317$ is needed for the normalization to $S^i(A,x,Q^2)$.
This form of the spatial dependence has a rather weak dependence on $s$ until
the nuclear surface is approached.  Note that
when $s \rightarrow 0$, $S^i_{\rm WS} < S^i$ in the shadowing and EMC regions 
while $S^i_{\rm WS}>S^i$ in the 
antishadowing region.

The actual spatial dependence of shadowing may be stronger if the
shadowing effect is not directly related to the nuclear matter density
distribution.  This can occur if the gluons are not well localized within the
nucleus.  One can alternatively assume that the shadowing is related to the 
nuclear thickness at the collision point, proportional to the
distance a parton from one nucleus travels through the other \cite{hijing}.
Therefore we also consider
\be
S^i_{\rm R}(A,x,Q^2,\vec{r},z) = \left\{ \begin{array}{ll}
 1 + N_{\rm R} (S^i(A,x,Q^2) - 1) \sqrt{1 - (r/R_A)^2} & 
\mbox{$r \leq R_A$} \\
 1                                                        &
\mbox{$r > R_A$} \end{array} \right.
 \, \, ,
\ee 
where $N_{\rm R}=1.449$ assures the normalization after the average over
$\rho(s)$.  Similarly,
when $s \rightarrow 0$, $S^i_{\rm R} < S^i$ in the shadowing and EMC regions 
while $S^i_{\rm R} > S^i$ in the 
antishadowing region.  The normalization is higher here because of the larger
region over which the suppression due to shadowing is reduced relative to
$S^i_{\rm WS}$.

We calculate the $c \overline c$ production cross sections in 
peripheral nuclear collisions with $S^i(A,x,Q^2) = 1$, $S_1$, and $S_2^i$.
As we will show, the shape of the inclusive charm quark $p_T$ distributions
are similar for $S_1$ and $S_2^i$.  Therefore, we model
the spatial dependence of $S_1$ only, according to eqs.\ (19) and (20).

\section{Correlation between E$_T$ and impact parameter}

Although the impact parameter is not directly measurable it can be
related to direct observables.  We discuss here the indirect measurement of the
impact parameter $b$ by means of the transverse energy $E_T$
\cite{us,eskola2}.  Here $E_T=\Sigma_i \sqrt{m_i^2+p_{Ti}^2}$, summed
over all detected particles in the event with masses $m_i$ and transverse
momenta $p_{Ti}$.  It is
also possible to infer the impact parameter by a measurement of the nuclear 
breakup since the beam remnants deposited in a zero
degree calorimeter are correlated with the impact parameter.  A
measure of the total
charged particle multiplicity, proportional to $E_T$, could be used to refine 
the impact parameter determination.

The transverse energy contains ``soft'' and ``hard'' components. The
``hard'' components arise from quark and gluon interactions above
momentum $p_0$, the scale above which perturbative QCD is assumed to be valid. 
Minijet production, calculated for $p_{T, {\rm jet}}> p_0
\sim 2$ GeV \cite{eskola3},
becomes an important contribution to the dynamics of the system in high-energy
nucleus-nucleus collisions.  The hard cross section, $\sigma_H^{pp}(p_0) = 
2\sigma_{\rm jet}$, twice the single LO minijet production cross section, can 
be calculated perturbatively.
``Soft'' processes with $p_T<p_0$ are not perturbatively calculable
yet they produce a
substantial fraction of the measured $E_T$ at high energies (and almost the
entire $E_T$ at CERN SPS energies).  These processes must be modeled 
phenomenologically.  We assume $\sigma_S^{pp} = \sigma_{\rm inelastic}^{pp}$,
the inelastic $pp$ scattering cross section.  Our calculation of the total
$E_T$ distribution follows Ref.\ \cite{eskola2}.

If the hard component is formed by independent parton-parton
collisions, then the 
average number of hard parton-parton collisions as a function of 
$b$, $N_{AA}^H(b)$, is
\begin{equation}
\overline N_{AA}^H (b) = T_{AA}(b)\sigma^{pp}_H (p_0) \, \, ,
\end{equation}
where $\sigma^{pp}_H (p_0) \sim 6.5$ mb at
RHIC \cite{eskola3} and $T_{AA}(b)$ is
the nuclear overlap function,
\begin{equation}
T_{AA} (b) = \int d^2r T_A(\vec{r}) T_A(\vec{b}-\vec{r})
\end{equation}
where the nuclear thickness function
is defined as $T_{A}(\vec{r}) = 
\int dz \rho_A(z, \vec r)$.  In Au+Au collisions at $b=0$, $T_{AA}=29$/mb 
\cite{overlap}.  
The $E_T$ distribution can be expressed as \cite{eskola2}
\begin{equation}
{d\sigma_H \over dE_T} = \int d^2b {\large \Sigma}_{N=1}^\infty 
{[\overline N_{AA}^H(b)]^N \over N!} \exp{[-\overline N_{AA}^H(b)]}
\int \prod_{i=1}^N dE_{Ti} {1 \over \sigma^{pp}_H} {d\sigma^{pp}_H \over 
dE_{Ti}} \delta(E_T - {\large \Sigma}_{i=1}^N E_{Ti}) \, \, .
\end{equation}
If $\overline N_{AA}^H$ is large, $d\sigma_H/dE_T$
can be approximated by the Gaussian \cite{eskola2}
\begin{equation}
{d\sigma_H \over dE_T} = \int d^2b {1 \over \sqrt{2\pi\sigma^2_H(b)}}
\exp{\bigg( \,- \, { (E_T-\overline E_{T \, H}^{AA}(b))^2 
\over 2\sigma^2_H(b)} \bigg)}, \end{equation}
where the mean $E_T$, $\overline E_{T \, H}^{AA}(b)$, and standard deviation, 
$\sigma_H(b)$, are proportional to the first and second $E_T$ moments of the
hard cross section,
\begin{eqnarray}
\overline E_{T \, H}^{AA}(b) & = & T_{AA}(b) \sigma^{pp}_H(p_0) 
\langle E_T \rangle^{pp}_H  \\
\sigma^2_H(b) & = & T_{AA}(b) \sigma^{pp}_H(p_0) \langle E_T^2 \rangle^{pp}_H 
\, \, .
\end{eqnarray}
In the rapidity interval $|y| \leq 0.5$, $\sigma^{pp}_H(p_0) \langle E_T 
\rangle^{pp}_H \approx 17 \, {\rm mb\ GeV}$ and $\sigma^{pp}_H(p_0) \langle 
E_T^2 \rangle^{pp}_H \approx 70 \, {\rm mb\ GeV}^2$ \cite{eskola3}.

At RHIC energies, the hard part does not dominate the soft
component, proportional to the number of nucleon-nucleon collisions,
\begin{equation}
\overline N_{AA}^S (b) = T_{AA}(b)\sigma^{pp}_S \, \, ,
\end{equation}
where $\sigma^{pp}_S \sim 30$ mb.  Since the soft component is almost
independent of the collision energy, we assume, as in Ref.\ \cite{eskola2},
that the hard and soft components are separable on the $pp$ level and thus
independent of each other at fixed $b$. Therefore the total $E_T$ distribution
is a convolution of the hard and soft components with total mean and
standard deviation 
\begin{eqnarray}
\overline E_T^{AA}(b) & = & T_{AA}(b) [\sigma^{pp}_H(p_0) 
\langle E_T \rangle^{pp}_H + \epsilon_0] \\
\sigma^2(b) & = & T_{AA}(b) [\sigma^{pp}_H(p_0) \langle E_T^2 \rangle^{pp}_H 
+ \epsilon_1] 
\end{eqnarray}
where $\epsilon_0$ and $\epsilon_1$ are taken from lower energy data and
adjusted to the same rapidity interval as the hard component, $|y| \leq 0.5$,
$\epsilon_0 = 15$ mb GeV, $\epsilon_1=50$ mb GeV$^2$ \cite{eskola2}.
Shadowing, which affects the hard component by reducing
the minijet cross section, is not included in these averages.  
Multiplying $\sigma^{pp}_H$ by a shadowing factor
modifies the $E_T$ distribution by less than
10\% \cite{eskola4}.  A correction has been included here.

Figure 2 shows the $E_T$ distribution (for $y<|0.5|$) for 100
GeV/nucleon Au+Au collisions for several different impact parameter 
intervals as well as the total cross section. Singling out a particular $E_T$ 
range can therefore select a rather
narrow distribution of impact parameters.  For example,
requiring $E_T< 300$ GeV selects almost exclusively events
with  $b>R_A$ while $E_T < 180$ GeV selects events with
$b > 1.2R_A$.

Good event purity can be obtained with even narrower selections.  For
example, 300 GeV $> E_T > 180$ GeV largely corresponds to $1.2R_A>b>R_A$.
An example of the purity can be seen in Fig. 3 which shows the range
of impact parameters at $E_T=200$ GeV. The distribution is centered at
$b=1.27R_A$ with a standard deviation $\sigma\sim0.05 R_A$. Approximately
90\% of the events fall into the range $1.15 R_A <b< 1.35 R_A$, narrow
enough to be an effective impact parameter selector. Thus at $E_T = 200$ GeV,
the impact parameter can be measured to within 10\%.  However, the statistical
accuracy depends on the average number of collisions, proportional to $E_T$,
so that $\sigma/b \approx 1/\sqrt{E_T}$.

For very small $E_T$, complications arise.  
The first concerns the transition from eq.\ (23) to
eq.\ (24) which is only valid if $\overline N_{AA}^H(b)$ is
large enough for the Poisson distribution 
to be approximated by a Gaussian.  For a small number of collisions,
eq.\ (24) overestimates the number of low $E_T$ events, even allowing
a finite probability for negative $E_T$ events. In practice, the
agreement is quite good even at $b=1.8R_A$, corresponding to
$T_{AA}=0.9$/mb, $N_{AA}^H=5.5$ and $E_T\approx30$~ GeV.  At 
significantly smaller
$E_T$ a correction is needed.  Further, the event by event
fluctuations are large when the collision number is small, increasing
the uncertainty in the impact parameter measurement.

At small $E_T$ the presence of charmed quarks will alter the
relationship between $E_T$ and impact parameter because a 
$c \overline c \rightarrow D \overline D$ pair
must have $E_T> 2m_D\approx
3.7$~GeV. Typical values are $E_T \sim 4-6$ GeV.  Thus when $E_T < 20$ GeV,
the relationship between $E_T$ and $b$ in charm events will be different.
This altered relationship can be studied in simulations to correct the data.

Finally, other types of interactions can contribute to charm
production at low $E_T$.  The largest identified charm contribution in
very peripheral collisions is photon-gluon fusion \cite{greiner,us2}.

Any real detector can only measure $E_T$ in a limited
rapidity interval.  For example, the calorimeter of the STAR detector at
RHIC will cover the range $-1<y<2$\cite{christie}. The 
acceptance can be compensated by appropriately modifying $\langle E_T 
\rangle^{pp}_H$, $\langle E_T^2 \rangle^{pp}_H$, 
$\epsilon_0$ and $\epsilon_1$, given here for $|y|<0.5$.  
The  accuracy scales roughly as
the square root of the observed energy. A large acceptance can also
extend the region of validity of eq.\ (24) to larger $b$. 

The non-central event selection technique to constrain the impact parameter
may be useful in other analyses of heavy
ion data.  At large impact parameters, only the outer portions
of the nuclei are involved but as the collision centrality increases,
the nuclear interior is more deeply probed.  Therefore the impact parameter
variation roughly corresponds to the portion of the nucleus involved in the 
interaction, and can thus be used to study the difference between the parton
constituents of the nuclear core and those near the surface.

\section{Results}

The best way to determine the gluon momentum fraction is to detect
both charm quarks.  Then $x_1$ and $x_2$ can be fixed exactly and the
shadowing mapped out.  The measurements are relatively easy to
interpret if $y = - \overline y$ since $x_1 = x_2$.  After first
discussing the general results when the kinematic variables are
integrated over, we show the $p_T$ distributions for the MRS
D$-^\prime$ and GRV HO parton densities assuming both the $c$ and
$\overline c$ are detected.  The low experimental efficiency for
detecting charm suggests that it is unlikely for both quarks to be
detected in an event.  Thus we subsequently discuss the feasibility of
the study if only one of the charm quarks is detected.

Figure 4 shows the $c \overline c$ production cross section as a
function of impact parameter for $b > R_A$ with $S=1$, $S_1$ and $S_2$
at RHIC\cite{RHIC}.  The cross sections were calculated by integrating
eq.\ (1) over the $c$ and $\overline c$ four-momenta.  The rates for
these non-central collisions are still quite large. Without shadowing,
for $b>1.2R_A$ the charm cross section is 2.9 b while for $b>1.8R_A$
it is still 200 mb.  At the RHIC Au+Au design luminosity,
$2\times10^{26} {\rm cm}^{-2} {\rm sec}^{-1}$ \cite{RHIC}, this
results in 6300 and 430 million $c \overline c$ pairs/year (3000
hours).  Thus these measurements will not be statistics limited, even
with the roughly 35\% reduction in cross section when shadowing is
included.

Figures 5 and 6 show the charm quark $p_T$ distributions in two
different impact parameter intervals, $b>1.2R_A$, roughly
corresponding to $E_T<180$ GeV in Fig. 2, and $b> 1.8 R_A$, for
several selected $c$ and $\overline c$ quark rapidities.  The results
with the MRS D$-^\prime$ and GRV HO parton densities are compared.  By
measuring charm quarks as a function of $p_T$ for a variety of
rapidities, different values of $x_1$ and $x_2$ are probed.  For
example, $p_T=0$, $y=\overline y=0$ corresponds to
$x_1=x_2=1.3\times10^{-2}$ while $p_T=0$, $y=2$ and $\overline y=-2$
corresponds to $x_1=x_2=5.1\times10^{-2}$.  At $p_T \approx 2.1$~GeV
$x_1$ and $x_2$ are doubled, moving into the antishadowing region for
$|y|=2$.  Thus varying $x_1$ and $x_2$ changes the relative strength
of the shadowing.  Calculations with $S=1$, $S_1$, $S_{1, {\rm WS}}$,
$S_{1, {\rm R}}$ and $S_2$ are shown in each case.

In every case considered, the unshadowed cross section is larger
than the shadowed cross sections.  
The total $c \overline c$ production cross
sections with $Q \propto nm_c$ differ only by 2\% in $pp$ collisions.
(Recall that $n=2$ for
MRSD$-^\prime$ and $n=1$ for GRV HO.) When the total cross
section is computed by integrating an inclusive cross section where
$Q \propto nm_T$, the difference increases to $\approx 6$\% due to the running
scale in the parton distributions and $\alpha_s$.  
The inclusive distributions reflect the low $x$ and $Q^2$ behavior 
of the parton distributions.
The MRS D$-^\prime$ gluon distributions are always decreasing as a
function of $p_T$.  However, the GRV HO gluon distributions are still
valence-like at low $Q$.  Thus for  $y = \overline y=0$ and
$p_T < 1.5$ GeV the gluon distribution continues to
increase, causing the observed $\approx 15$\% difference between the $S=1$
distributions at $p_T \approx 0$ in Figs. 5(a) and (d).  At larger rapidity and
$x$, such as in Figs. 5(c) and (f), the difference is reduced to $\approx 8$\%.

The shadowing functions affect the charm $p_T$ distributions
differently for the MRS D$-^\prime$ and GRV HO parton distributions
because of the difference in the scale $Q^2$.  In general $S_2^G$
increases more rapidly with $x$ than $S_1$ between the shadowing and
antishadowing regions.  With the MRS D$-^\prime$ parton distributions,
at $p_T \approx 0$, $S_1 \approx S_2^G$ for $Q \approx 2m_c$.  As
$p_T$ increases, $S_2^G > S_1$ due to the evolution of $S_2$.
Therefore when $p_T \approx 1$ GeV, the $p_T$ distribution with $S_2$
will be $\approx 10$\% larger than the distribution with $S_1$.  This
continues to hold as $p_T$ rises, as shown in Figs. 5(a), (b) and (c).
The GRV HO case is different because of the lower scale.  There, the
evolution of $S_2$ with $Q^2$ does not begin until $Q_0 = 2$ GeV,
corresponding to $p_T \approx 1.5$ GeV. For $p_T < 1.5$\ GeV, $S_1 >
S_2^G$.  At $p_T \approx 2$ GeV for $y = \overline y = 0$, the
evolution of $S_2^G$ causes the situation to be reversed and $S_2^G >
S_1$, as can be seen by inspection of Fig. 5(d).  At larger
rapidities, the larger slope of $S_2^G$ in the shadowing region cause
the switch between $S_1$ and $S_2$ dominance to occur at lower values
of $p_T$, even before the evolution of $S_2^G$ begins, since $x$ is
larger at small $p_T$ and large $y$.

Including spatial dependence in $S_1$ increases the cross section
toward the $S=1$ value at high $b$ where the nuclear density is low.
The cross section is now larger because the lower density near the
nuclear surface reduces the shadowing.  As the impact parameter rises,
the tails of the density distributions are probed and the shadowed
cross sections approach the $S=1$ result.  This happens relatively
slowly, especially for $S_{1, {\rm WS}}$, since the density is nearly
constant except within $d$ of the surface.  The shadowing is thus
almost constant except near the nuclear surface. For gold, $d=0.535$
fm while $b=1.8R_A$, the lower bound on the impact parameter in Fig.\
6, corresponds to collisions within 1.2 fm of the surface so that some
collisions occur below the surface layer in at least one nucleus.  In
both Figs.\ 5 and 6, $S_{1, {\rm R}} > S_{1, {\rm WS}}$ because the
dependence on the nuclear thickness (albeit for a spherical nucleus)
decreases the effects of shadowing already at small $r$ while $S_{1,
{\rm WS}}$ is almost constant.  The effect is more apparent for larger
impact parameters.  When $b>1.2 R_A$, both spatial forms increase the
cross section about 15\% over $S_1$.  For $b > 2R_A$ the spatial
results are approximately halfway between the cross sections with
$S=1$ and $S=S_1$.  The similarity of results between the two spatial
parameterizations suggests that the parton localization measurements
may not be too hard to interpret.

Thus measurements of charm quark production at large impact parameters
probe the nuclear surface where shadowing effects are greatly reduced,
and, for extremely peripheral collisions, the limit of independent
$pp$ collisions is regained.  As the collisions become more central,
the charm quark production rate should begin to deviate from the naive
expectation from superimposed $pp$ collisions.  By measuring charm
production as a function of impact parameter, it is possible to watch
the shadowing turn on with the rate of increase providing a measure of
parton localization in the nucleus.

So far we have assumed that both
the $c$ and $\overline c$ quarks are detected.  Given the low efficiency
for detecting charm quarks, either by their semileptonic decays or by
reconstruction of specific final states, it is worth considering what
can be learned if only one of the quarks is detected.
Fig.\ 7 shows the rapidity distribution of the $\overline c$ quark, 
assuming that the
$c$ quark is detected at $y=0$ and $p_T=0$ assuming $S=1$, $S_1$, $S_2$,
$S_{1, {\rm WS}}$ and $S_{1, 
{\rm R}}$.  Kinematically, this situation corresponds to charm pair invariant
mass $M^2 = 2m_c^2 (1 + \cosh \overline y)$ so that increasing $\overline y$ 
corresponds to increasing phase space along with increasing
invariant mass. The cross section increases until $y \approx \pm1$ where 
$M \approx 3.4$ GeV and decreases with larger $M$, typical for invariant
mass distributions\cite{HPC}.
Fig. 8 shows the single charm $p_T$ distribution 
at $y=0$\ integrated over $\overline y$
for $b>1.8R_A$.  The results are similar to the case when 
both quarks are detected.
Although some information is lost if only a single quark is detected,
the trends remain the same as those seen in Fig.\ 6.  
Therefore it should still be possible to 
extract the shadowing information from the data.

\section{Discussion}

If the charmed quark rapidity and momentum can be measured over a broad
range of impact parameters, the gluon momentum
distribution and its spatial/density dependence can be measured.  However,
there are a number of difficulties involved in relating these calculations to
measurements.  Charm is normally detected either via its
semileptonic decays or through reconstruction of selected decay
modes.  While the detection of leptons from semileptonic decays is
fairly straightforward, the lepton $p_T$ and $y$ differ from that of
the parent hadron.  The parent hadron distribution can also differ slightly 
from that of the
initially produced quark although the hadronic environment reduces this
effect \cite{VBH}.  While this momentum shift does not create any fundamental
problems, it adds another intermediate step which must be correctly
modelled.  Fully reconstructed charm decays such as $D^{*+}\rightarrow
D^0\pi^+\rightarrow (K^-\pi^+)\ \pi^+$ could allow for a full reconstruction
of the meson direction, reducing the uncertainty in the determination of
the charmed quark $p_T$ and $y$.  However, the small branching ratios and
low efficiency for detecting
these decays probably preclude the useful detection of both
charmed quarks in a pair.

In addition to gold, RHIC will accelerate a variety of lighter nuclei.
The surface layer is a larger fraction of the nuclear radius in
lighter nuclei.  In this case, the Woods-Saxon and square root spatial
dependencies should more closely match over the full range of impact
parameters.  Since RHIC is also a $pA$ collider, the gluon
localization could in principle be probed for an individual nucleus.
However, for $pA$, the number of collisions is small enough for the
Gaussian approximation to break down, rendering the $E_T$ to $b$
correlation problematic.  The $A$ dependence of charm production at
various impact parameters can in any case provide an additional handle
on interplay between shadowing and its spatial dependence.  For $pA$,
dileptons can also be used to probe gluon shadowing\cite{ziwei}.

At LHC, similar calculations can be made for $c\overline c$ and
$b\overline b$ production.  The higher energy
implies that the charm and bottom pairs will be produced at much lower $x$,
increasing the importance of shadowing and further reducing the 
production cross sections.  Thus the sensitivity of the cross section to the
spatial dependence will be enhanced.

\section{Conclusions}

We have calculated charmed quark production
in non-central Au+Au collisions for several different structure functions
and assumptions about nuclear shadowing. 

Shadowing reduces the charm production cross section up
to 35\%.  However, when the spatial dependence of shadowing is included, the
effect is decreased.  By measuring the
charmed quark production rates as a function of impact parameter, it
is possible to study the effect of shadowing and its localization within the
nucleus.  This spatial dependence provides an indication of the gluon 
recombination distance scale.

The correlation between impact parameter and transverse energy has been used
to fix $b$.  We have shown that the impact parameter determination is reliable
to within a 10\% statistical uncertainty on an event-by-event basis for
$b \approx 1.2R_A$. 

\section{Acknowledgements}

V.E. and A.K. would like to thank the LBNL Relativistic Nuclear Collisions 
group for their hospitality and  M. Strikhanov for discussions and support.
We also thank K.J. Eskola for providing the shadowing routines and for
discussions.  This work was supported in part by the Director,
Office of Energy Research, Division of Nuclear Physics of the Office of High
Energy and Nuclear Physics of the U. S. Department of Energy under Contract
Number DE-AC03-76SF0098.

\begin{figure}
\vspace*{6.5 in}
\hspace*{.45 in}
\includegraphics{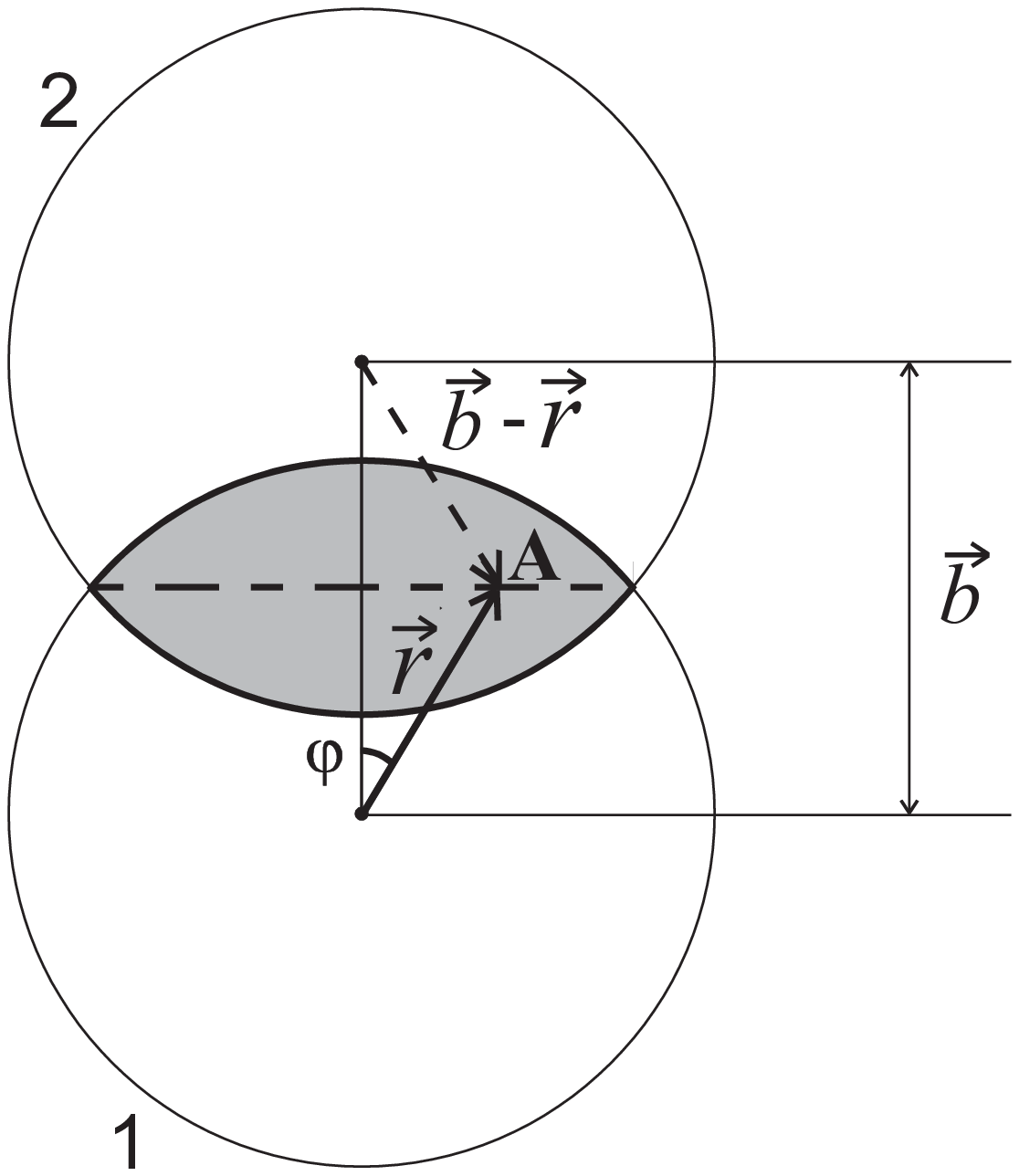}
\caption{The collision geometry of nuclear collisions in the plane transverse
to the beam.  The parton-parton collision point is indicated by $A$ and 
$b$ is the impact parameter.}
\end{figure} 

\begin{figure}
\centerline{\psfig{figure=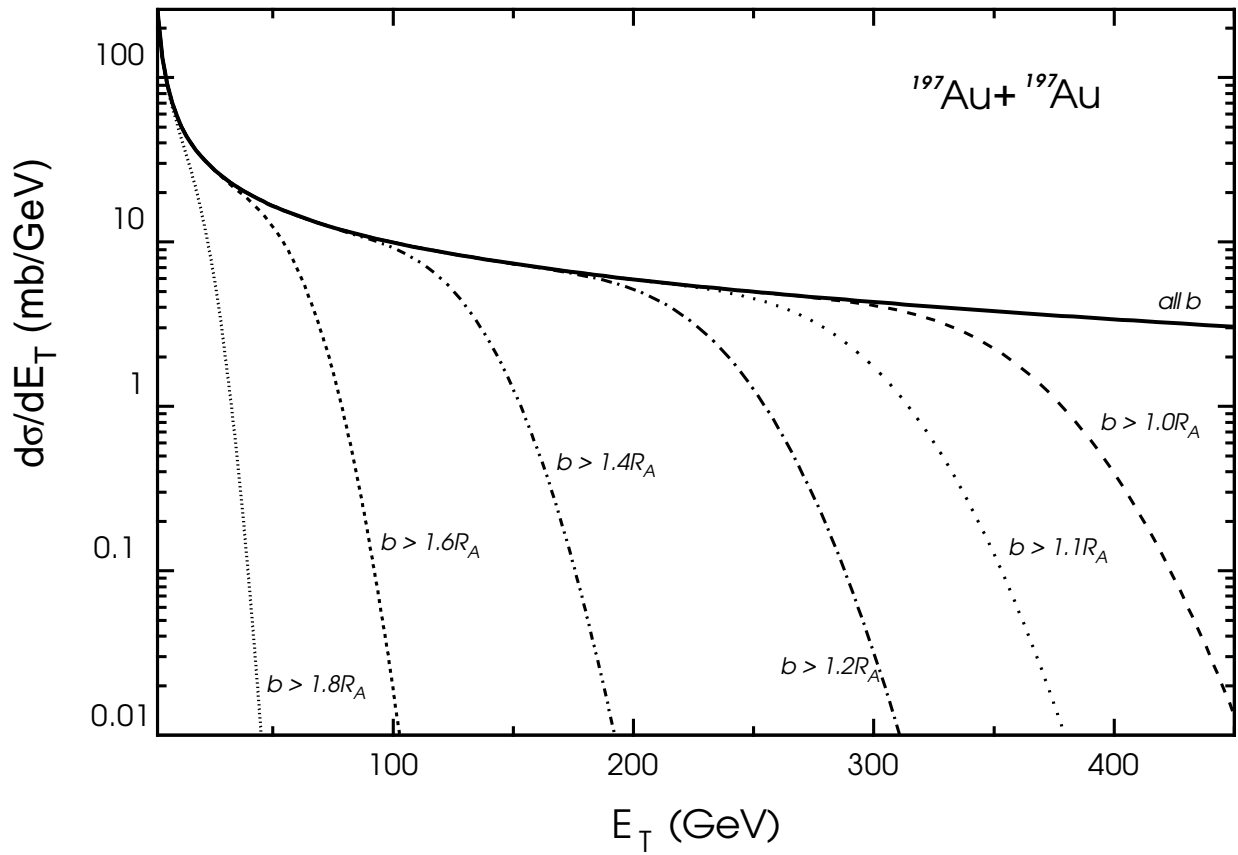,rheight=6. in,width=7.0 in}} 
\caption{Cross section as a function of $E_T$, for a selection
of impact parameters ranges.}
\end{figure} 

\begin{figure}
\centerline{\psfig{figure=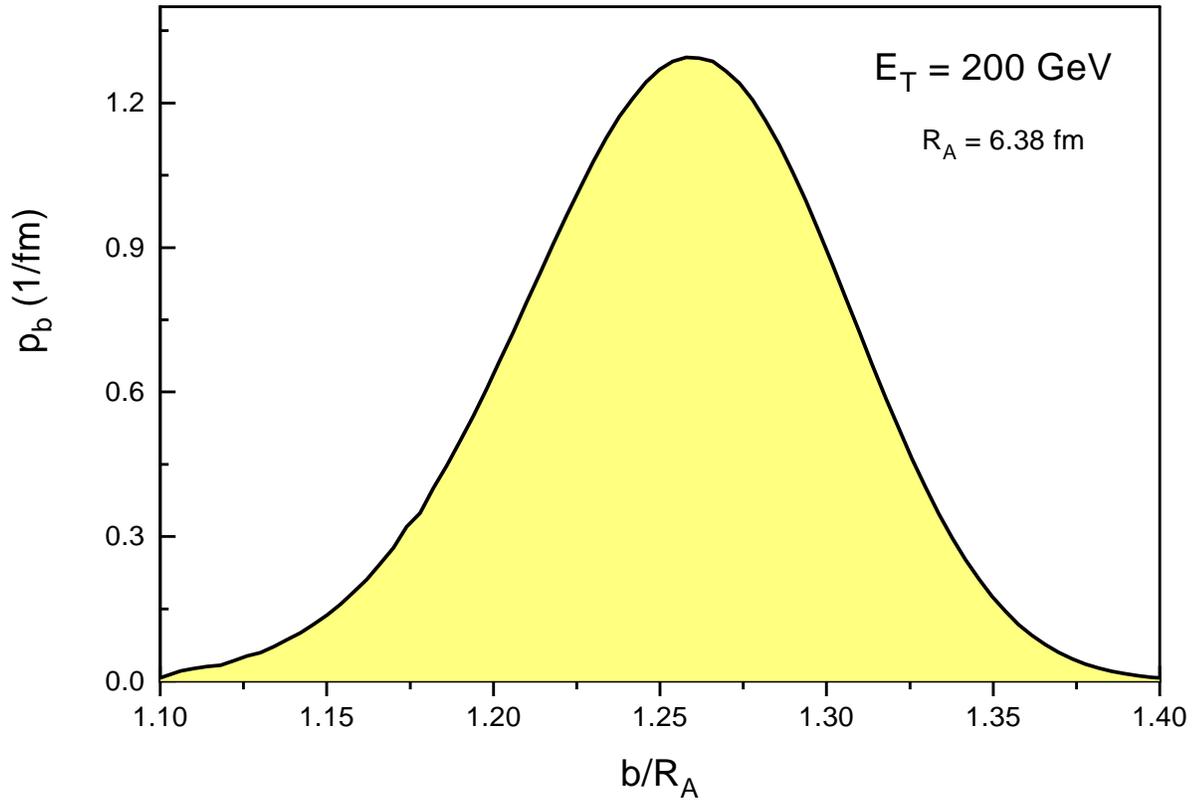,rheight=5.5 in,width=6.5 in}} 
\caption{Distribution of impact parameter for events
with $E_T=200$ GeV.}
\end{figure} 

\begin{figure}
\centerline{\psfig{figure=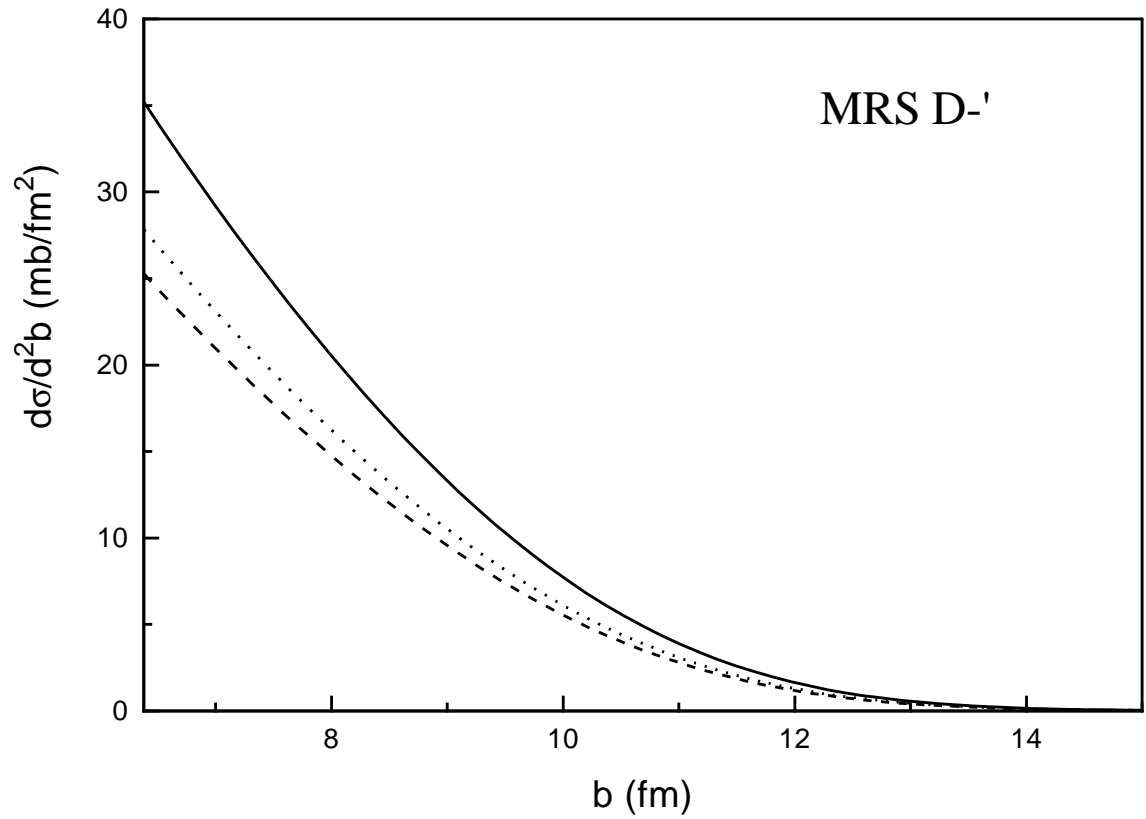,rheight=5.5 in,width=6.5 in}} 
\caption{Charm production cross section as a function of $b$ for the
MRS D$-^\prime$ parton densities, with $S=1$ (solid line)
and with two nuclear shadowing parameterizations $S_1$ (dashes) and $S_2$
(dots).}
\end{figure} 

\begin{figure}
\centerline{\psfig{figure=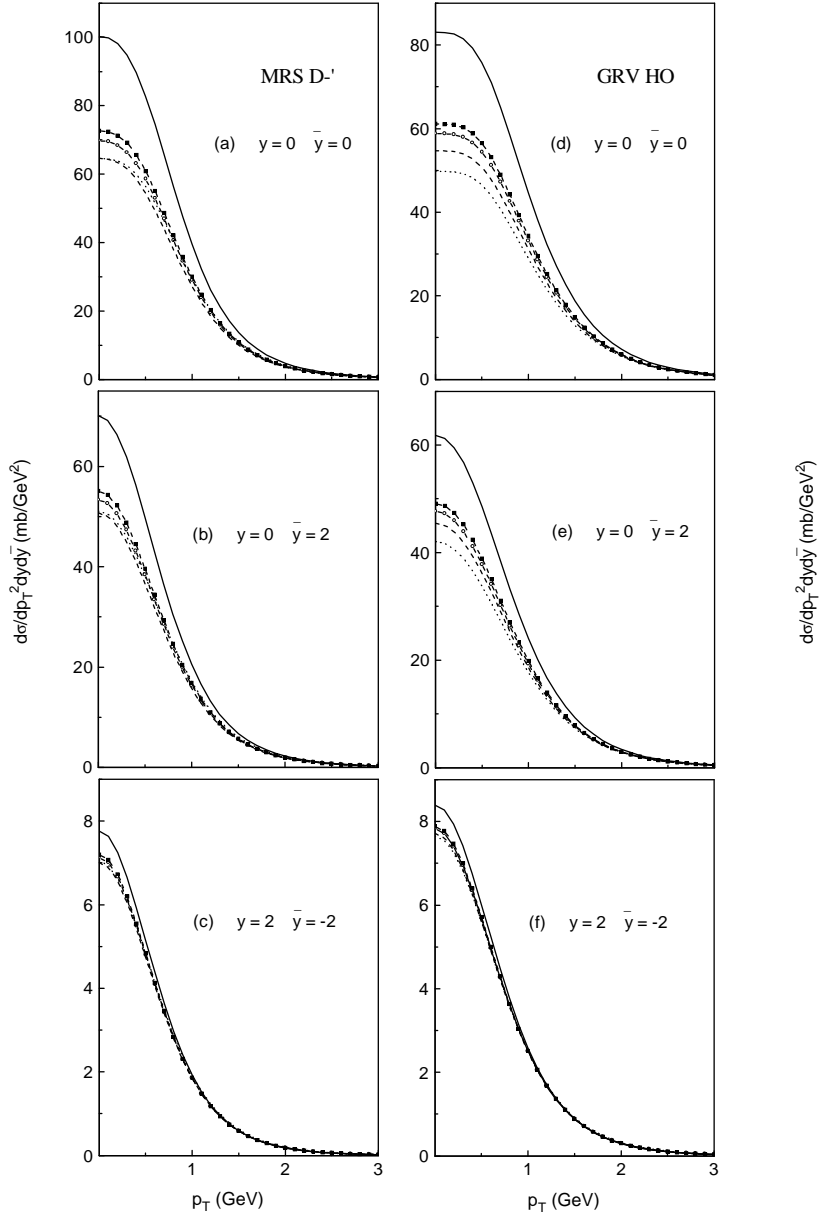,rheight=6.35 in,width=5.5 in}}
\vskip .5 in 
\caption{The $p_T$ distribution of $c \overline c$ pairs for the MRS
D$-^\prime$ (a), (b) and (c) and GRV HO (d), (e) and (f) parton
densities.  We select events with $b> 1.2R_A$ and 3 sets of $c$ and
$\overline c$ quark rapidities: $y=0$, $\overline y=0$ in (a) and (d);
$y=0$, $\overline y = 2$ in (b) and (e); $y=2$, $\overline y = -2$ in
(c) and (f).  The solid curves are for $S=1$.  The spatially
independent shadowing results are given by the dashed ($S_1$) and
dotted ($S_2$) curves.  The effect of the spatial dependence on $S_1$
is also shown.  The dashed curve with the filled squares shows the
result with $S_{1, {\rm R}}$ and the dashed curve with the open
circles gives the result with $S_{1, {\rm WS}}$.  In (a), (b), (d) and
(e) the $S_1$ and $S_{1,WS}$ curves overlap.}
\end{figure} 

\begin{figure}
\centerline{\psfig{figure=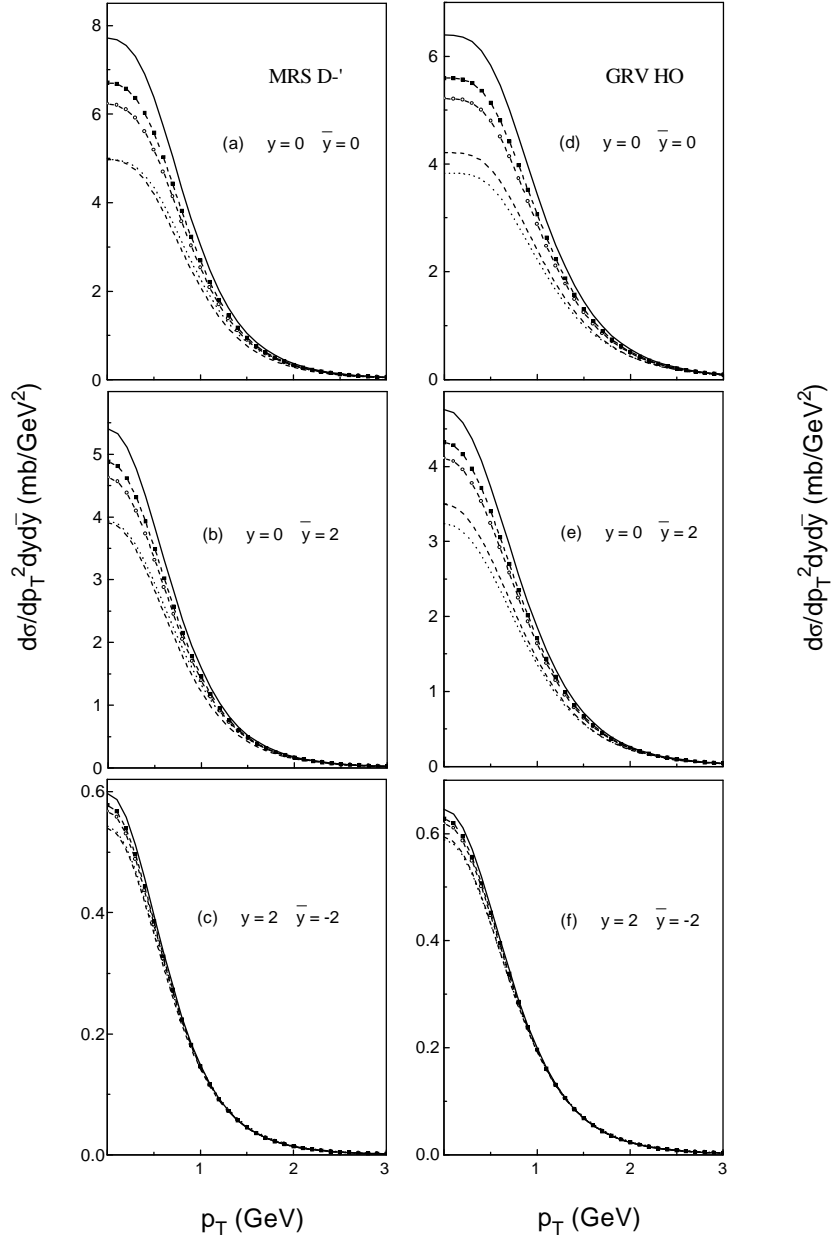,rheight=6.35 in,width=5.5 in}} 
\vskip .5 in
\caption{The same as in Fig.\ 5 but with $b>1.8R_A$.}
\end{figure} 

\begin{figure}
\centerline{\psfig{figure=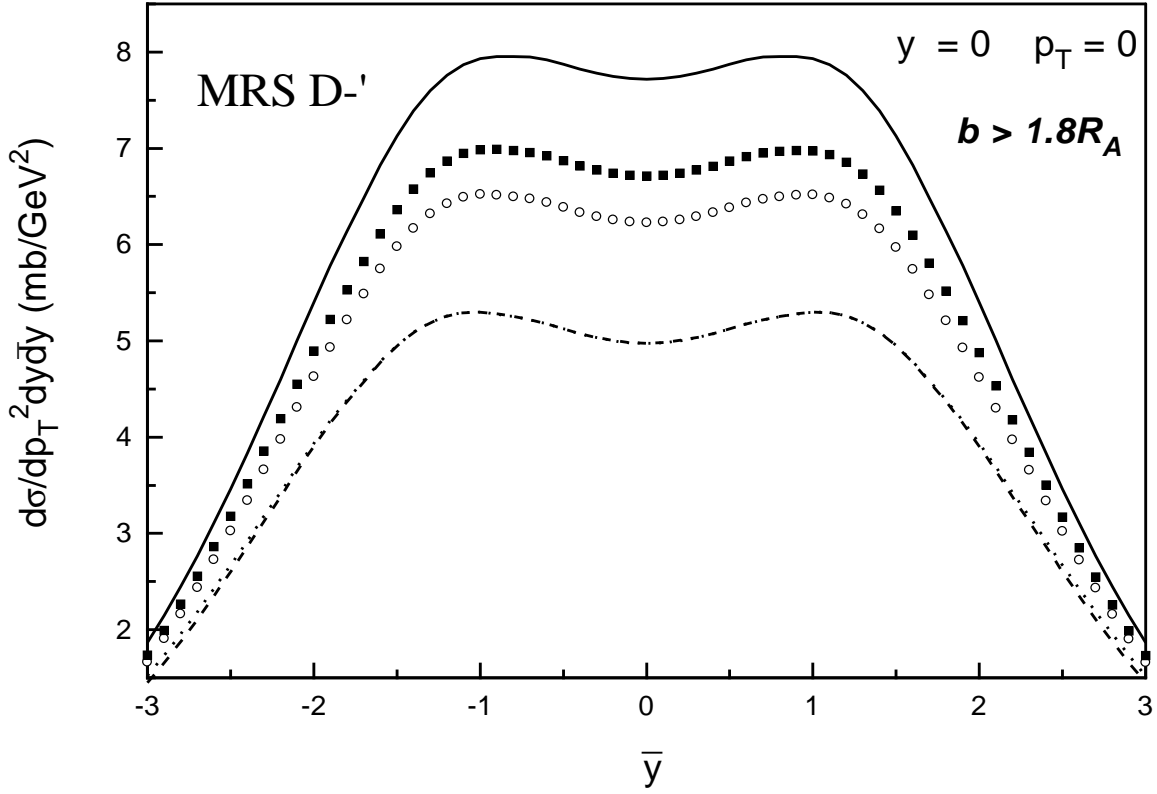,rheight=5.5 in,width=6.5 in}} 
\caption{The $\overline c$ rapidity distribution for $p_T=0$ and the charm
quark is produced at $y=0$.  The solid
curve is with $S=1$.  The spatially independent shadowing results are
given in the dashed and dotted curves for $S_1$ and $S_2$ respectively.
The effect of the spatial dependence on $S_1$ is also shown.  The 
filled squares shows the result with $S_{1, {\rm R}}$ and the
open circles gives the result with $S_{1, {\rm WS}}$.}
\end{figure} 

\begin{figure}
\centerline{\psfig{figure=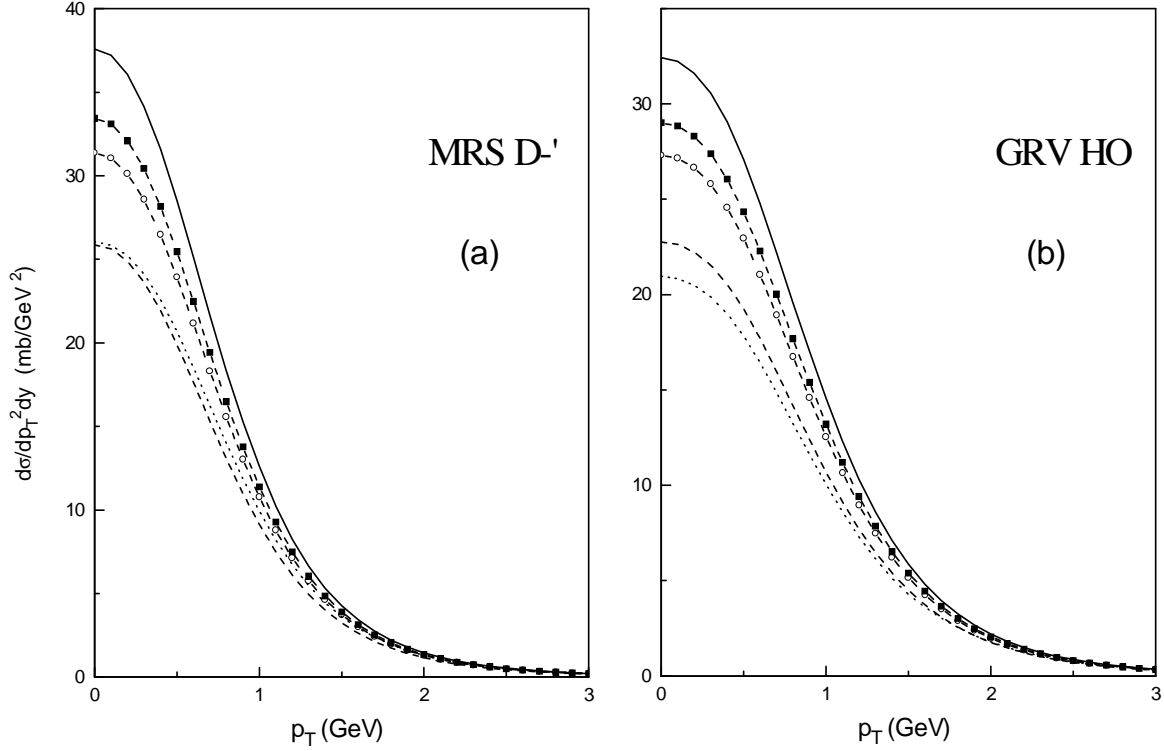,rheight=7 in,width=6.5 in}} 
\caption{The $p_T$ distribution for single charm quarks with $y=0$ for 
MRS D$-^\prime$ (a) and GRV HO (b) parton 
densities.  We have selected events with $b> 1.8R_A$.  The solid
curves are with $S=1$.  The spatially independent shadowing results are
given in the dashed and dotted curves for $S_1$ and $S_2$ respectively.
The effect of the spatial dependence on $S_1$ is also shown.  The dashed curve
with the filled squares shows the result with $S_{1, {\rm R}}$ and the
dashed curve with the open circles gives the result with $S_{1, {\rm WS}}$.}
\end{figure} 
\vskip 1in
\end{document}